\documentclass{PoS}

\title{Supersymmetry Searches with Multiple $b$-jets at CMS}

\ShortTitle{CMS SUSY with $b$-jets}

\author{\speaker{Keith A. Ulmer}\\
        University of Colorado, Boulder\\
        E-mail: \email{keith.ulmer@colorado.edu} \\
	\\
        on behalf of the CMS Collaboration\\        
	}

\abstract{Recent results from CMS are reviewed for searches for supersymmetry in final states
with multiple bottom quark jets. Results are based on the full 2012 CMS dataset consisting of
19.5 fb$^{-1}$ collected at a center-of-mass energy of $\sqrt{s} = 8$ TeV. In particular, searches
for final states with multiple $b$-jets and one or two leptons are presented. These final states
are of special interest in the context of the search for third generation squarks in
gluino or sbottom cascade decays, as predicted by natural supersymmetry.}

\FullConference{The European Physical Society Conference on High Energy Physics -EPS-HEP2013\\
		18-24 July 2013\\
		Stockholm, Sweden}

\begin{document}

\section{Introduction}

After the discovery
of the Higgs Boson~\cite{CMShiggs,ATLAShiggs}, one of the most pressing questions to
address at the LHC is what mechanism suppresses the quantum divergences that would
appear as corrections to the Higgs mass. A natural solution to this hierarchy problem
has justifiably become a key focus of studies at the LHC. We present here the results
from two such searches at the Compact Muon Solinoid (CMS)~\cite{CMS} experiment. In each,
multiple bottom ($b$) quarks are considered for final states resulting from the decay 
products of third generation squarks. First, a search in single lepton (electron or muon)
events is presented, followed by a search in events with two same-charge leptons. In both
searches the results are consistent with the expectations from standard model backgrounds
and limits are set on a variety of new physics models, including natural supersymmetric
scenarios.

\section{Single lepton search}
A search for physics beyond the standard model is performed in events with large
hadronic activity defined by the sum of the transverse momentum of hadronic jets in the
event (HT), transverse missing energy, and one isolated electron or muon~\cite{RA4}. 
The search
is motivated by models of new physics, including supersymmetry, that involve
strong production processes and cascade decays producing many jets and missing
momentum from unobserved weakly interacting particles. At least one of the jets is required
to be tagged as from a bottom quark.

Two independent methods are used to search for new physics and to 
estimate the size of the standard model background. In 
the first method, the missing energy spectrum is predicted by observing the charged lepton
spectrum in data. When considering charged leptons and
missing energy from leptonically decaying $W$ bosons, the two spectra should be similar, up 
to acceptance, resolution and $W$
polarization effects, which are taken into account. The dominate background for this 
search is $t\bar{t}$ events with one leptonically decaying $W$ boson, thus allowing this
``Lepton Spectrum'' method to provide the main data-driven background estimate. Sub-dominant
background contributions from mis-reconstructed dilepton events are predicted using the kinematics 
of a control sample of fully reconstructed dilepton events.

The second method makes use of the angle between the charged lepton and the $W$ momentum vectors,
$\Delta\phi(W,l)$.
The angle is expected to be small for standard model $t\bar{t}$ events where the lepton comes
from the $W$, while it can take on large values for new physics contributions where the missing
energy and the lepton are not as correlated. Backgrounds are further suppressed by
requiring the leptonic mass scale of the event~\cite{RA4} to be large. The search requires
$\Delta\phi(W,l) > 1$, and uses the data control region $\Delta\phi(W,l) < 1$ to estimate the
standard model background to the search. The ratio of the signal to control region in 
$\Delta\phi(W,l)$ is obtained from a data control region with exactly one $b$-tag, while the search
is performed in the regions with 2 or $\geq 3$ $b$-tags.

\begin{figure}[htb]
\centering
\includegraphics[width=0.48\linewidth]{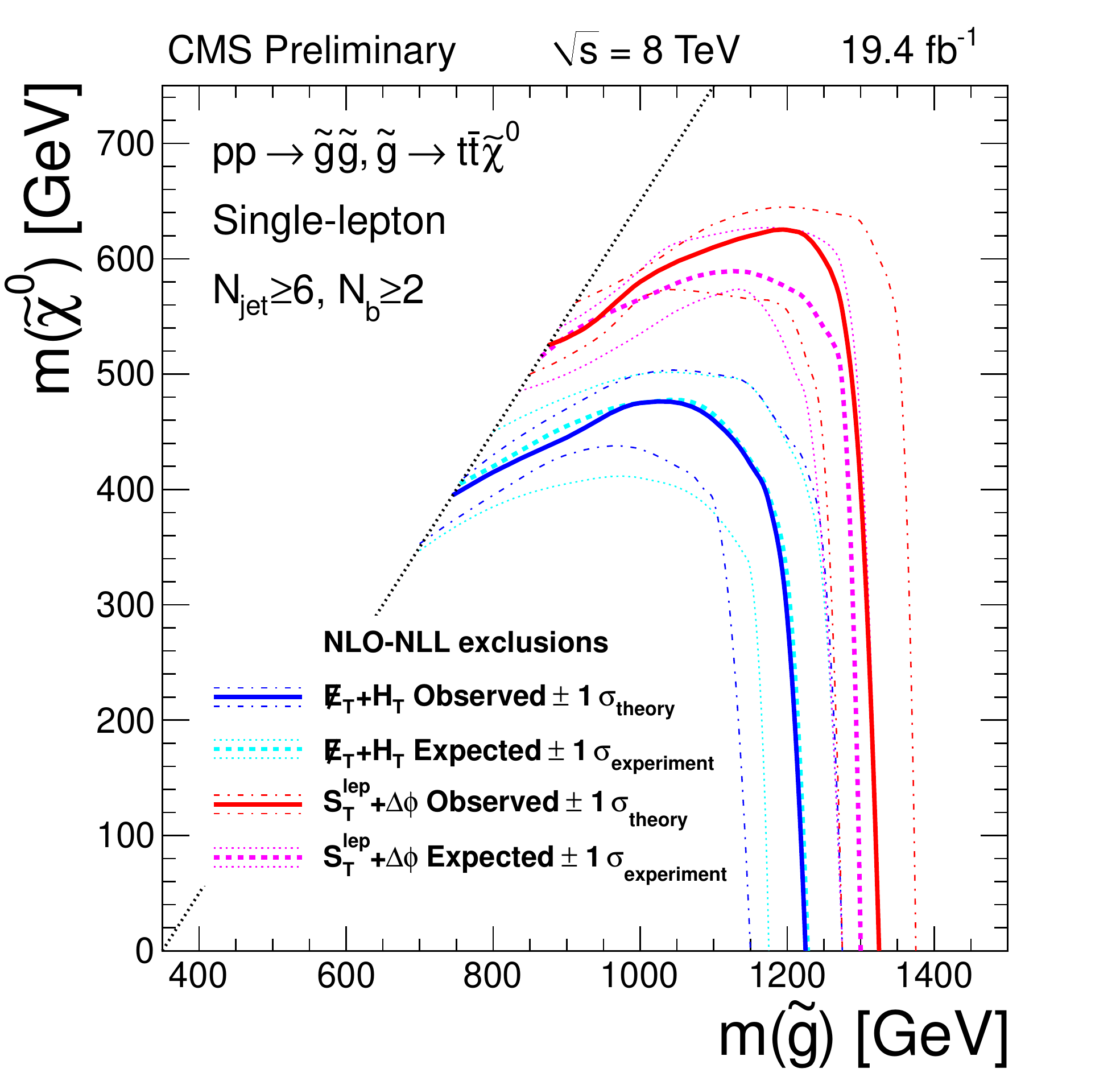}
\caption{Limits from the two single lepton search methods for gluino pair production
with each gluino decaying to $t\bar{t}\tilde{\chi}^0$.}
\label{fig:RA4}
\end{figure}

No excess of events over the predicted standard model background is observed with either method.
These results are interpreted as limits in the simplified model~\cite{SMS} of gluino pair production
with each gluino decaying to two top quarks and the lightest supersymmetric particle (LSP). This
scenario is the dominant decay of the gluino when all squarks are heavier than the gluino, with
the stop squark as the lightest squark. Results from the search with both methods are shown in
Fig.~\ref{fig:RA4}. The most sensitive limits come from the $\Delta\phi(W,l)$ search, which
excludes gluinos up to masses $\sim$ 1.3 TeV for the case of light LSP masses.

\section{Dilepton search}
Standard model events with two isolated, like-charge leptons are rare. New physics 
processes, however, can enhance the rate of these events, making the signature a good
place to search for physics beyond the standard model. In this search, a pair of 
isolated electrons or muons is required where both leptons must have the same charge~\cite{RA5}.
The same-sign dilepton signature can be produced by a broad range of new physics scenarios.
To be sensitive to potential contributions in different regions of phase space, the
search is divided into non-overlapping regions based on missing transverse energy, HT, number of jets,
number of $b$-jets, and the transverse momentum of the leptons.

In each signal region, the expected contribution from standard model events is estimated. The main
sources of background events are the following: ``fake'' leptons that originate from non-prompt
sources, such as $b$-hadron decays; rare standard model processes with two true like-sign, isolated
leptons, such as $t\bar{t}W$ events; and opposite-charge dilepton events where the charge of one of
the leptons is misidentified. The ``fake'' lepton background is measured with a data-driven technique
extrapolating from a looser isolation control region to the signal region. The rare background is 
estimated from simulation. The wrong charge background is measured in data from dilepton events
reconstructed with the same charge with a dilepton mass that peaks at the $Z$ mass
indicating a true opposite charge lepton pair.

\begin{figure}[htb]
\centering
\includegraphics[width=0.48\linewidth]{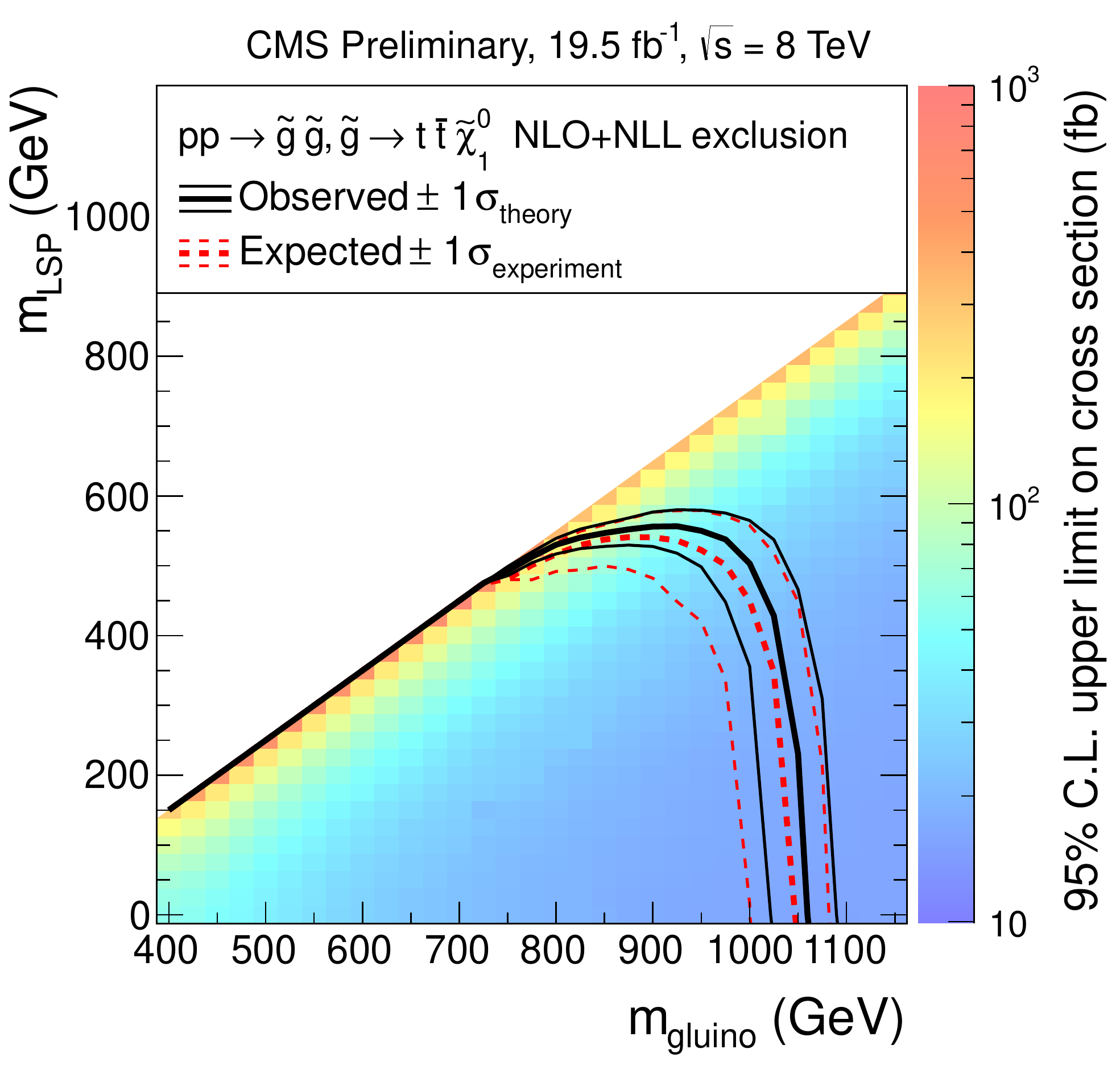}
\includegraphics[width=0.48\linewidth]{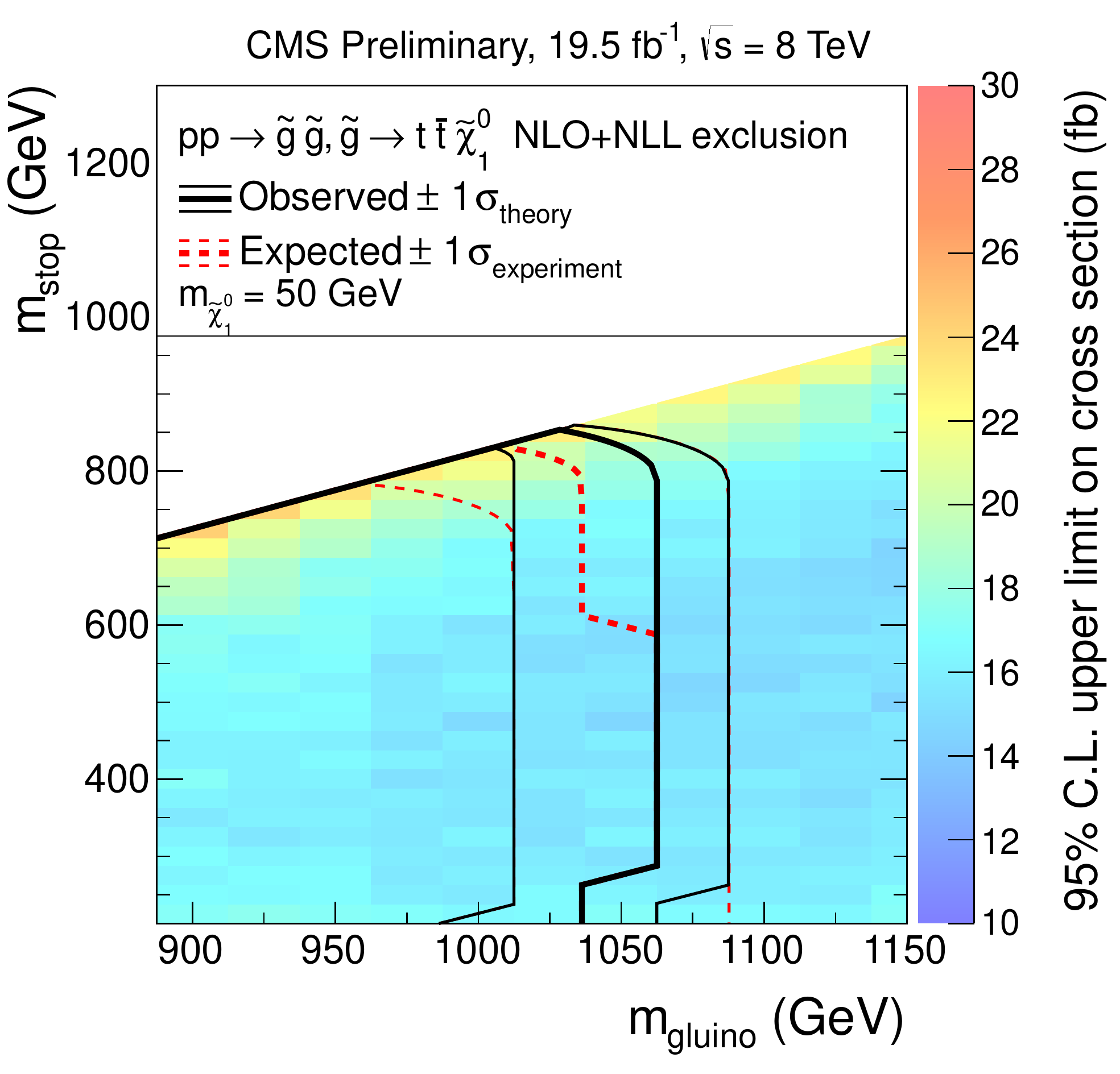}
\caption{Limits from the same-sign dilepton search for gluino pair production
with each gluino decaying to $t\bar{t}\tilde{\chi}^0$ (left) and each gluino decaying to
$t\tilde{t}$, with $\tilde{t}\rightarrow t\tilde{\chi}^0$ (right).}
\label{fig:RA5-glu}
\end{figure}

\begin{figure}[htb]
\centering
\includegraphics[width=0.48\linewidth]{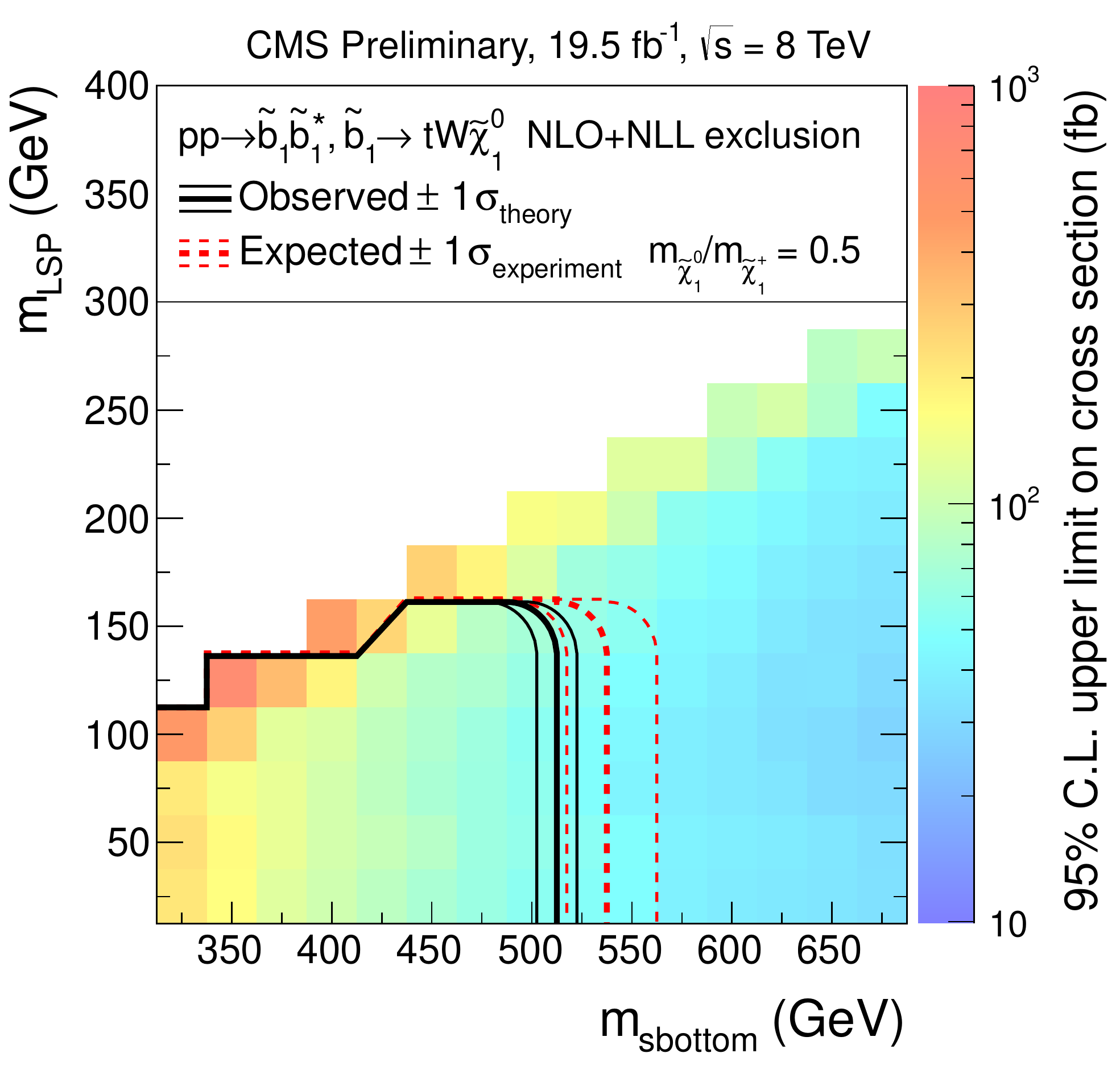}
\includegraphics[width=0.48\linewidth]{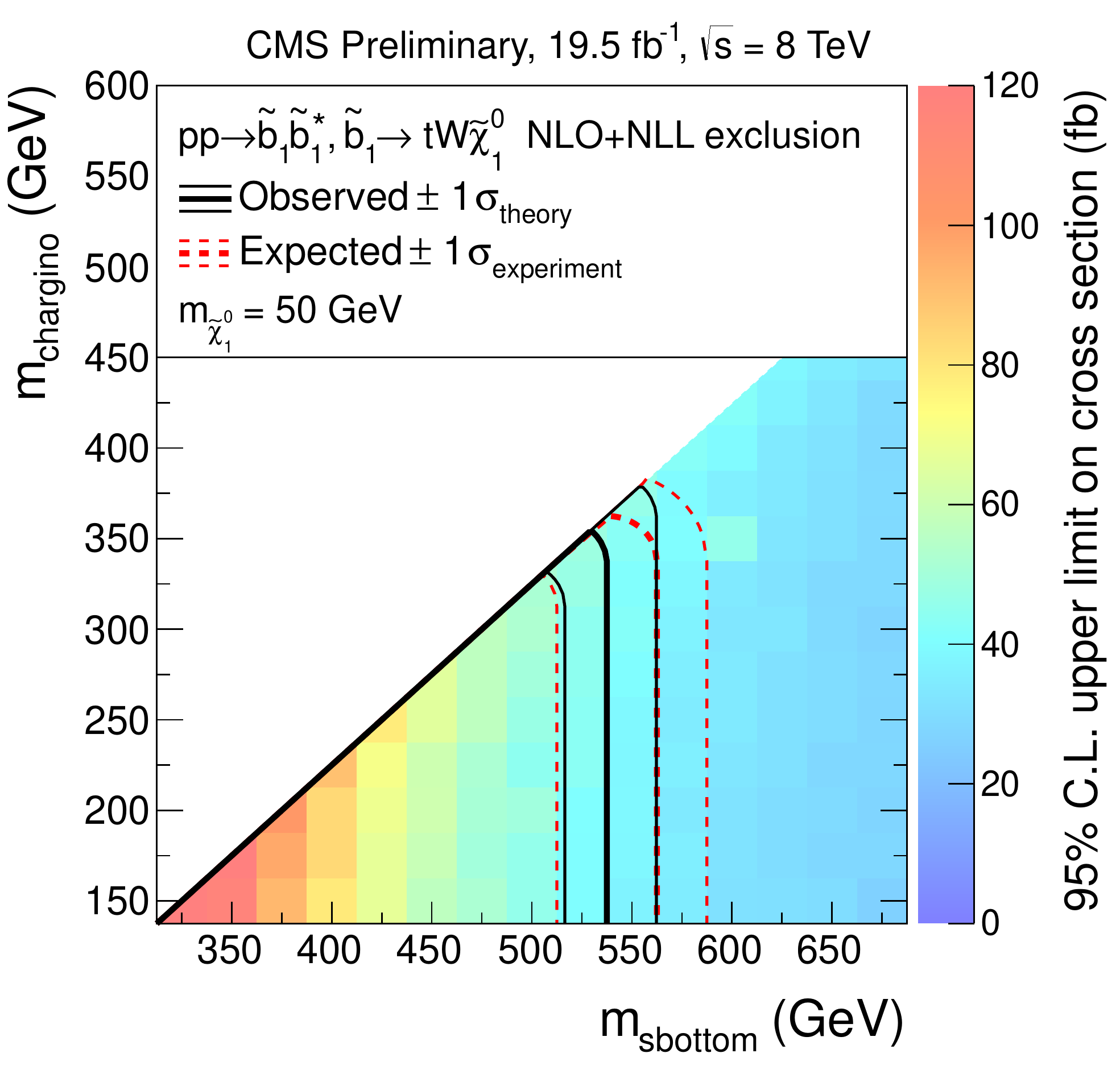}
\caption{Limits from the same-sign dilepton search for sbottom pair production
with each sbottom decaying to $tW\tilde{\chi}^0$ for two different mass splitting
scenarios as indicated in the plots.}
\label{fig:RA5-sb}
\end{figure}

No excess is observed in any of the signal search regions. Confidence limits at the level of 95$\%$ are
set for a variety of signal models, including gluino-mediated stop production, and sbottom production.
Figure~\ref{fig:RA5-glu} left shows the results for the same gluino-mediated stop production model as
was used for the single lepton search. The right plot of Fig.~\ref{fig:RA5-glu} shows results for a 
similar model where the stop is instead allowed to be on-shell. Gluinos with masses up to 
$\sim$ 1000 GeV are ruled out for low LSP or stop mass in these scenarios. The search is
also sensitive to pair production of sbottoms, with each sbottom decaying to top, $W$, LSP, as shown
in Fig.~\ref{fig:RA5-sb}. The left and right plots show two scenarios as labeled in the plots for
different mass splittings between the relevant supersymmetric particles. Limits up to $\sim$ 500 GeV
for the sbottom mass are obtained in these scenarios.

\section{Conclusions}
\begin{figure}[htb]
\centering
\includegraphics[width=0.48\linewidth]{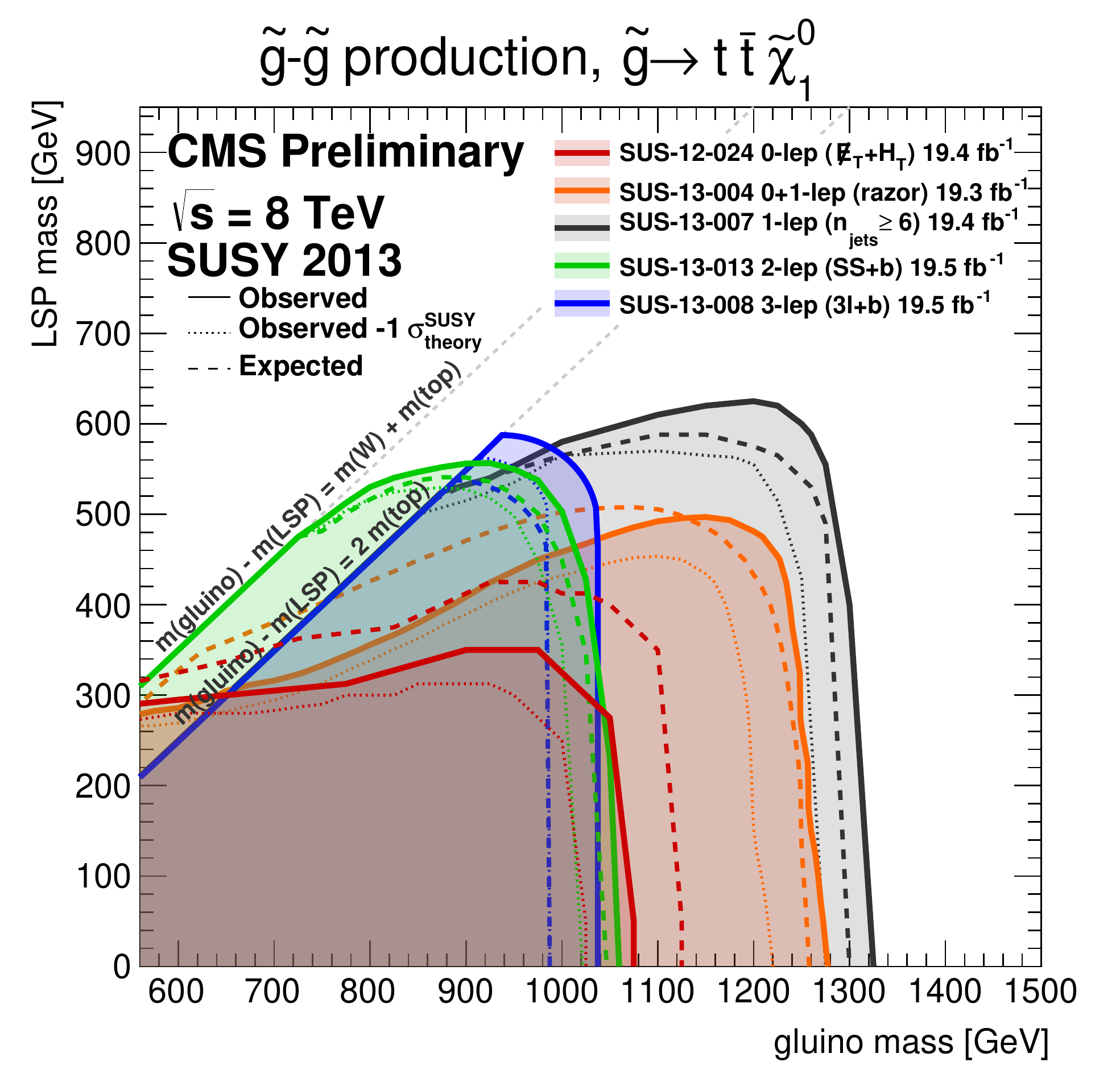}
\caption{Summary of CMS results for gluino pair production
with each gluino decaying to $t\bar{t}\tilde{\chi}^0$.}
\label{fig:summary}
\end{figure}

Results from CMS searches for events with multiple $b$-tagged jets and
either exactly one lepton, or a pair of opposite-charge leptons were presented.
These searches
are part of a rich CMS program searching for final states with 4 $W$ bosons and
multiple $b$-tags, a topology that can be searched for with a number of leptons
ranging from 0-4. Figure~\ref{fig:summary} shows a summary of CMS search limits
for the off-shell gluino-mediated stop scenario, including the results presented here, plus
a fully hadronic search~\cite{RA2b}, a search combining hadronic and single lepton
final states~\cite{razor}, and a search with $\geq 3$ leptons~\cite{multilep}.
Thus far, no evidence of supersymmetry has been observed.

\end{document}